\documentclass[epj]{webofc}
\usepackage[utf8]{inputenc}
\usepackage[varg]{txfonts}   
\usepackage{booktabs}
\usepackage{xcolor}
\definecolor{darkred}{rgb}{0.4,0.0,0.0}
\definecolor{darkgreen}{rgb}{0.0,0.4,0.0}
\definecolor{darkblue}{rgb}{0.0,0.0,0.4}
\usepackage[bookmarks,linktocpage,colorlinks,
    linkcolor = darkred,
    urlcolor  = darkblue,
    citecolor = darkgreen]{hyperref}
\newcommand{\beq}{\begin{equation}}
\newcommand{\eeq}{\end{equation}}

\newcommand{\cA}{{\cal A}}
\newcommand{\cAb}{{\overline{\cal A}}}

\newcommand{\cF}{{\cal F}}
\newcommand{\cFb}{{\overline{\cal F}}}
\newcommand{\cD}{{\cal D}}
\newcommand{\cDb}{{\overline{\cal D}}}
\newcommand{\cQ}{{\cal Q}}

\newcommand{\cU}{{\cal U}}

\newcommand{\cN}{{\cal N}}
\newcommand{\cUb}{{\overline{\cal U}}}

\newcommand{\Tr}{{\rm Tr\;}}

\newcommand{\phib}{{\overline{\phi}}}

\def\bec{\begin{center}}
\def\eec{\end{center}}
\def\beq{\begin{equation}}
\def\eeq{\end{equation}}
\def\bea{\begin{eqnarray}}
\def\eea{\end{eqnarray}}
\usepackage{subfigure}
\wocname{EPJ Web of Conferences}
\woctitle{Lattice2017}
\begin{document}
%
\selectlanguage{english}
\title{%
Two dimensional super QCD on a lattice
}
\author{%
\firstname{Simon} \lastname{Catterall}\inst{1}\fnsep\thanks{Speaker, \email{smcatter@syr.edu}} \and
\firstname{Aarti} \lastname{Veernala}\inst{2}
}
\institute{%
Physics Department, Syracuse University, Syracuse NY13244
\and
Fermilab, PO Box 500, Batavia, IL 60510
}
\abstract{%
  We construct a lattice theory with one exact supersymmetry which consists of fields transforming in
  both the adjoint and fundamental representations of a U(Nc) gauge group. In addition
  to gluons and gluinos, the theory contains Nf flavors
  of fermion in the fundamental representation along with their scalar partners and is invariant under
  a global U(Nf) flavor symmetry. The lattice action 
  contains an additional Fayet-Iliopoulos term which can be used to generate a scalar potential.
  We perform numerical simulations that corroborate the theoretical
  expectation that supersymmetry is spontaneously broken for Nf<Nc. 
}
\maketitle
\section{Introduction}\label{intro}

A great deal of progress has occurred over the last decade
in the construction of
supersymmetric lattice theories. The key realization has been that
in certain theories with extended supersymmetry it is possible to find linear combination(s)
of the original supercharges that satisfy a nilpotent subalgebra of the full
supersymmetry algebra which can be transferred intact to the lattice. The nature of this subalgebra is
most clearly seen after the continuum theory is {\it twisted} - that is the theory is recast in a new set of
variables that exposes a particular  subgroup of the original global symmetries. 
This subgroup is the diagonal subgroup of the Lorentz and flavor symmetries - which are hence twisted
or locked together. Details of these lattice constructions can be
found in \cite{Sugino:2004qd,Catterall:2004np,physrep}.~\footnote{The same lattice theories can be obtained using orbifold methods and indeed supersymmetric lattice actions  for Yang-Mills theories were first constructed using
this technique \cite{Cohen:2003xe,Cohen:2003qw,Kaplan:2005ta,Damgaard} and the connection between twisting and orbifold methods forged in \cite{Unsal:2006qp}}
Initially the focus was on lattice actions that target pure super Yang-Mills theories in the continuum limit with a great deal of
effort being devoted to $\cN=4$  super Yang-Mills
\cite{latsusy-1,latsusy-2,Catterall:2012yq,Catterall:2011pd,Catterall:2013roa}.

In \cite{Matsuura} and \cite{SuginoQuiver} these formulations
were extended using orbifold methods to the case of  theories incorporating fermions transforming in the fundamental representation of the gauge group. These methods yield  quiver gauge
theories containing fields that transform as bi-fundamentals under a product gauge group $U(N_c)\times U(N_f)$. 

In this work we derive these quiver actions by generalizing the existing twisted SYM lattice constructions in dimension $d+1$ to achieve a quiver theory and 
show how, in a certain limit, these reduce to a class of super QCD theory in dimension $d$.  In practice we
consider only the two dimensional case although it should be clear how this
approach will generalize. We also discuss how to add
a lattice supersymmetric Fayet-Illipoulos term to the action which will play a role in mediating
spontaneous supersymmetry breaking.  We have performed numerical simulations  to test these
ideas and our results
show strong evidence for spontaneous susy breaking for $N_f<Nc$ as expected on general theoretical grounds.

\section{Eight supercharge SYM on a 3d lattice}\label{sec-1}

To target ${\cal N}=(2,2)$ SQCD in two dimensions we start from the continuum eight supercharge ($\cQ=8$) theory in three dimensions. Conventionally this
is written as
\beq
S= \int d^3x\, \left(F_{ab}^2+\left(D_a B^I\right)^2+\left[B^I,B^J\right]^2\right)+\;{\rm fermions}\eeq
Here the lowercase indices are spacetime indices while the uppercase indices denote flavor.  All fields are in the adjoint representation of a $U(N_c)$ gauge group $X=\sum_{a=1}^{N^2} X_a T_a$ and we will adopt an antihermitian basis for the
generators $T_a$.  After twisting this can be rewritten
in terms of fields which are completely antisymmetric tensors in spacetime under the twisted SO(3) group given by
\beq
SO_{\rm twist}(3)={\rm diag}\left[SO_{\rm rot}(3)\times SO_{\rm flavor}(3)\right]\eeq
The original two Dirac fermions
reappear in the twisted theory as the components of a K\"{a}hler-Dirac field
$\Psi=\left( \eta, \psi_{a}, \chi_{ab}, \theta_{abc} \right)$.
The bosonic sector of the twisted theory comprises  a complexified gauge field $\cA_a=A_a+iB_a$ containing the 
original gauge field $A_a$ and an additional vector field $B_a$. This additional field
contains the original three scalars which, being vectors under the R symmetry,
transform as the components of a vector field after twisting.
The corresponding action $S=S_{\rm exact}+S_{\rm closed}$ becomes
\bea
S_{\rm exact} &=& \frac{1}{g^2} \; \cQ \Lambda = \frac{1}{g^2} \; \cQ \int d^3x {\rm Tr} \left[ \chi_{ab}(x)\cF_{ab}(x) + \eta(x)\left[\cDb_{a},\cD_{a}\right] + \frac{1}{2}\eta(x)d(x) \right], 
\label{quiverActionQexact}\\
S_{\rm closed}&=& - \; \frac{1}{g^2} \int d^3x  {\rm Tr} \left[\theta_{abc}(x) \cDb_{[c}\chi_{ab]}(x) \right].
\label{quiverActionQclosed}
\eea 
 where $\cD_{a}$ and $\cDb_{a}$ are the continuum covariant derivatives defined in terms of the complexified gauge fields as $\cD_{a} = \partial_{a} + \cA_{a}$ and $\cDb_{a} = \partial_{a} + \cAb_{a}$.
The action of the scalar supersymmetry on the twisted fields is given by
\begin{eqnarray}
\cQ \cA_a &=&  \psi_a\nonumber\\
\cQ \cAb_a &=& 0\nonumber\\
\cQ \psi_a &=&  0 \nonumber \\
\cQ \chi_{ab} &=&  -\cFb_{ab}\nonumber\\
\cQ \eta &=&  d   \nonumber\\
\cQ \theta_{abc} &=& 0
\end{eqnarray} 
Notice that we have included an auxiliary field $d(x)$ that allows the algebra to be off-shell nilpotent $\cQ^2=0$. 
This feature
then guarantees that $S_{\rm exact}$ is supersymmetric. 
The $\cQ$-invariance of $S_{\rm closed}$ follows from the Bianchi identity\footnote{Note that it is also possible to write the 3d action completely in terms of an $\cQ$-exact form without a $\cQ$-closed term by employing
an additional auxiliary field $B_{abc}$} \\
\beq
\epsilon_{abc}\cDb_{c}\cFb_{ab} = 0.
\label{bianchi}
\eeq 
To discretize this theory we place all fields on the links of a lattice. This 3d lattice consists of the usual hypercubic vectors plus additional face and body links. In detail
these assignments are
\medskip
\begin{center}
\begin{tabular}{c|c}\hline
continuum field & lattice link\\
$\cA_a(x)$ & $x\to x+\hat{a}$\\
$\cAb_a(x)$ & $x+\hat{a}\to x$\\
$\psi_a(x)$ & $x\to x+\hat{a}$\\
$\chi_{ab}$ & $x+\hat{a}+\hat{b} \to x$\\
$\eta(x)$ & $x\to x$\\
$d(x)$ & $x\to x$\\
$\theta_{abc}$ & $x\to x+\hat{a}+\hat{b}+\hat{c}$ \\
\hline
\end{tabular}
\end{center}\bigskip
The lattice gauge field will be denoted $\cU_\mu(x)$ in the following discussion. Notice that the orientation of a given fermion
link field is determined by the even/odd character of its corresponding continuum antisymmetric form. 
The link character of a field determines its transformation properties under lattice gauge transformations eg. $\cU_a(x)\to G(x)\cU_a(x)G^\dagger(x+\hat{a})$. 
To complete the construction of  the lattice action it is necessary to replace continuum covariant derivatives by appropriate gauged lattice difference operators. The necessary
prescription was described in \cite{physrep},~\cite{twist2orb}. It is essentially determined by the simultaneous requirements that 
the lattice difference agree with the continuum derivative as the lattice spacing is sent to zero and that it yields expressions that transform
as the appropriate link field under lattice gauge transformations. 
For example the continuum derivative $D_a\psi_b$ becomes
\beq
\cD^{(+)}_a\psi_b(x)=\cU_a(x)\psi_b(x+\hat{a})-\psi_b(x)\cU_a(x+\hat{b})\eeq
This prescription yields a set of link paths which, when contracted with the link field $\chi_{ab}(x)$, yields a closed loop whose trace is gauge invariant:
\beq
{\rm Tr}\;\left[\chi_{ab}(x)\left(\cU_a(x)\psi_b(x+\hat{a})-\psi_b(x)\cU_a(x+\hat{b})\right)\right]\label{typical}
\eeq
The definition of the field tensor $\cF_{ab}=\cD^{(+)}_a \cU_b(x)$ follows similarly and guarantees an exact
lattice Bianchi identity.

\section{Quiver lattice gauge theories}
Consider a typical term occurring in the 3d twisted action eg. eqn.~\ref{typical} . Let's do a gauge transformation
\beq
{\rm Tr}\,\left[G(x+a+b)\chi_{ab}(x)G^\dagger(x)G(x)\psi_a(x)G^\dagger(x+a)G(x+a)\cU_a(x)G^\dagger(x+ab+b)\right]\eeq
where $G(x)$ denotes an element of $U(N_c)$. 
It should be clear that this term remains invariant when we promote $\chi$ to be a bi-fundamental field transforming as
$\chi(x)\to H(x+a+b)\chi(x)G^\dagger(x)$  where
$H(x)$ takes values in another group $U(N_f)$ provided we simultaneously transform $\cU_b(x)$ as $\cU_b(x)\to G(x)\cU_b(x)H^\dagger(x+b)$. In this
case we can no longer think of $\cU_b$ as a gauge field but instead we treat it as a bi-fundamental scalar connecting
2d lattice theories with distinct  gauge groups $U(N_c)$ and $U(N_f)$.  The resultant quiver lattice gauge theory clearly
retains the original exact lattice supersymmetry of the 3d model.

Consider a lattice whose extent in the b-direction comprises just two 2d slices.  Let us take $b=3$. Furthermore we shall assume free boundary conditions in the 3-direction
so that these two slices are connected by just a single set of links in the 3-direction - those running from
$x_3=0$ to $x_3=1$. Denoting directions on the 2d slices by Greek indices
$\mu,\nu = 1,2 $ the fields living entirely on these lattices are given by
\bea
N_{c} \; \; \; &:& \; \; \; \Psi(x) = \left( \eta, \psi_{\mu}, \chi_{\mu\nu} \right), \; \; \; \cU_{\mu} = I_{N_{c}} + \cA_{\mu},\qquad d
\label{Nc-ferm} \\
N_{f} \; \; \; &:& \; \; \; \hat{\Psi} (\overline{x})= \left( \hat{\eta}, \hat{\psi}_{\mu}, \hat{\chi}_{\mu\nu} \right), \; \; \; \hat{\cU}_{\mu} = I_{N_{f}} + \hat{\cA},_{\mu}, \qquad \hat{d}
\label{Nf-ferm}
\eea 
In these expressions $x(\overline{x})$ denotes the coordinates on the $N_{c}(N_{f})$ lattice and $1_{N_{c}(N_{f})}$ denote the $N_{c}(N_{f}) \times N_{c}(N_{f})$ unit matrix respectively.
Now consider fields that live on the links between the $N_c$ and $N_f$ lattice. These must necessarily transform as bi-fundamentals under $U(N_c)\times U(N_f)$.
We have,
\bea
N_{c} \; \times \; N_{f}  \; \; \; &:& \; \; \;\Psi_{\text{bi-fund}}(x,\overline{x}) = \left( \psi_{3}, \chi_{\mu3}, \theta_{\mu\nu3} \right) = \left( \lambda, \lambda_{\mu}, \lambda_{\mu\nu}\right),\qquad \phi
\label{bifund}
\eea  The second equality in the above equation is 
a mere change of variables and corresponds to labeling fields according to their two dimensional character.
The complete field content of this model is summarized in the table below: \\
\begin{center}
\begin{tabular}{ c | c | c}
$N_{c}$-lattice & Bi-fundamental fields & $N_{f}$-lattice \\ 
$x$ & $(x,\overline{x})$ , $(\overline{x},x)$ & $\overline{x}$ \\ \hline
& & \\
$\cA_{\mu}(x)$ & $\phi(x,\overline{x})$   & $\hat{\cA_{\mu}}(\overline{x})$\\ 
$\eta(x)$ & $ \lambda(x,\overline{x})$  & $\hat{\eta}(\overline{x})$\\ 
$\psi_{\mu}(x)$ & $\lambda_{\mu}(\overline{x}+\mu,x)$ & $\hat{\psi}_{\mu}(\overline{x})$\\
$\chi_{\mu\nu}(x)$ &  $\lambda_{\mu\nu}(x,\overline{x}+\mu+\nu)$  & $\hat{\chi}_{\mu\nu}(\overline{x})$\\  
& & \\ \hline
\end{tabular} 
\end{center} 
As before we will write $G(x)$ as a group element belonging to $U(N_{c})$ and $H(x)$ to $U(N_{f})$. The lattice
gauge transformations for the bi-fundamental fields are as follows: 
\begin{eqnarray}
\phi(x) &\rightarrow& G(x)\phi(x)H^{\dagger}(\overline{x})\nonumber\\
\lambda(x) &\rightarrow& G(x)\lambda(x)H^\dagger(\overline{x})\nonumber\\
\lambda_{\mu}(x) &\rightarrow& H(\overline{x}+\mu)\lambda_\mu(x)G^{\dagger}(x)\nonumber\\
\lambda_{\mu \nu}(x) &\rightarrow &G(x)\lambda_{\mu \nu}(x)H^{\dagger}(\overline{x}+\mu+\nu)
\label{gaugetrans}
\end{eqnarray}
For example the 3d term given in eqn.~\ref{typical} on the $N_c$ time slice ($x_3=0$) yields 
\begin{eqnarray}
\chi_{a3}\cD_a\psi_3 &\to& \lambda_{\mu}(x)\left(\cU_\mu(x)\lambda(x+\mu)-\lambda(x)\cU_\mu(x)\right)\\
\chi_{3a}\cD_3\psi_a &\to& \lambda_{\mu}(x)\left(\phi(x)\psi_\mu(x)-\psi_\mu(x)\phi(x+\mu)\right)
\end{eqnarray}
We see that we obtain kinetic terms for the bi-fundamental fields and Yukawa interactions mediated by the bi-fundamental scalars between
the adjoint and bi-fundamental fermions. Furthermore by construction these terms
are invariant under the the generalized gauge transformations given in eqn.~\ref{gaugetrans}. 

Thus, the above construction lends us a consistent lattice supersymmetric quiver gauge theory containing both adjoint and bi-fundamental fields 
transforming under a product  $U(N_c)\times U(N_f)$ gauge group. At this point we can send the gauge coupling on the
$N_f$ lattice to zero thereby driving the $N_f$ gauge links to unity up to gauge transformations and their superpartner fermions to zero. The original
$U(N_f)$ gauge  symmetry now becomes a global flavor symmetry and the theory can be effectively truncated to a single 2d time slice.
On this slice the bi-fundamental fields yield 
$N_f$ fermions and their scalar superpartners transforming in the fundamental of $U(N_c)$ - super QCD.
The SUSY transformations for the remaining adjoint and fundamental fields are:  
\begin{center}
\begin{tabular}{ c | c}
Adjoint Fields & Fundamental fields \\ \hline \\
$ \cQ \cA_{\mu} =  \psi_{\mu}$  & $ \cQ \phi = \lambda $ \\
$ \cQ \cAb_{\mu} = 0$ & $ \cQ \phib = 0 $ \\
$ \cQ \psi_{\mu} =  0 $ & $\cQ \lambda = 0$ \\
$ \cQ \chi_{\mu\nu} =  -\cFb_{\mu\nu}$ & $\cQ \lambda_{\mu} = -\cDb_{\mu} \phib$  \\
$ \cQ \eta = d$ & $\cQ \lambda_{\mu\nu} = 0 $ \\
&  \\
\end{tabular} 
\end{center} 
At this point we have the freedom to add to the action one further supersymmetric and gauge invariant term - namely $r\sum_x \Tr d(x)=r \cQ \sum_x \Tr \eta$. This is a Fayet-Iliopoulos 
term. Its presence changes the equation of motion for the auxiliary field
\beq
d(x)=\cDb^{(-)}_\mu\cU_\mu(x)+\phi^a(x)\phib^a(x)-rI_{N_c}
\label{dFI} \eeq
with  $I_{N_c}$ a $N_c\times N_c$ unit matrix.  After integration over $d$ the Fayet-Iliopoulos term
yields a scalar potential term which will play a crucial role in determining whether the system can undergo spontaneous supersymmetry
breaking.
\beq
V=\frac{1}{2}\sum_{x}  {\rm Tr}  \left( \sum_f \phi^f(x)\phib^f(x) - \rm {rI_{Nc}} \right)^{2}
\eeq

In practice we have also included the following soft SUSY breaking mass term in the action $S_{\rm soft}$ which lifts the flat directions associated with the
adjoint scalars buried in the complex gauge fields $\cA_\mu$.
\beq
S_{\rm soft} = \mu^{2}\left[ \frac{1}{N_c} \Tr \left( \cUb_{\mu}\cU_{\mu} \right) - 1 \right]^{2}.
\label{actionSoft}
\eeq

\section{Vacuum Structure and SUSY Breaking Scenarios}

Supersymmetry will be spontaneously broken if the vacuum has non-zero energy $V>0$.
Consider the
case where $N_f<N_c$.
Using $SU(N_c)$ transformations one can diagonalize the $N_c\times N_c$ matrix $\phi\phib$. In general it will have
$N_f$ non-zero  real, positive eigenvalues and $N_c-N_f$ zero eigenvalues.  This immediately implies that there
is no configuration of the fields $\phi$ where the potential is zero. Indeed the minimum of the potential will
have energy $r^2(N_c-N_f)$,and corresponds to a situation where $N_f$ scalars develop vacuum expectation values breaking the gauge group to $U(N_c-N_f)$. The situation when $N_f\ge N_c$ is qualitatively different;
now the rank of $\phi\phib$ is at least $N_c$ and a zero energy vacuum configuration is possible. In such a situation
$N_c$ scalars pick up vacuum expectation values and the gauge symmetry is completely broken. \\

\section{Numerical Results}
\begin{figure}
\begin{center}
\includegraphics[height=50mm]{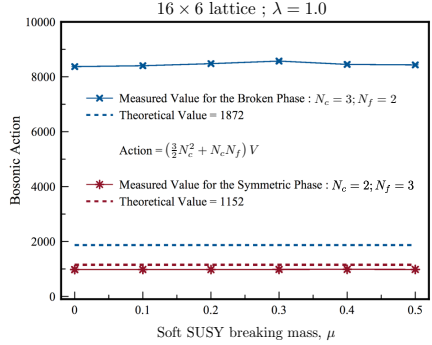}
\caption{Normalized bosonic action vs soft breaking coupling $\mu$ for $\lambda = 1.0$ for a 16x6 lattice}
\label{baction}
\end{center}
\end{figure}

In this section, we contrast results from simulations
with $N_f=2$, $N_c=3$ corresponding to the predicted
susy breaking scenario with results from simulations with $N_f=3$, $N_c=2$ - the susy preserving case. We ran our simulations for three different values of the `t~Hooft coupling, $\lambda=0.5,1.0$ and $1.5$ and observed the same qualitative behavior for the different values of $\lambda$. 
The results presented here  correspond to $\lambda=1.0$. The FI parameter, r, is a free parameter and is set to 1.0 for the rest of the discussion. \\
As a first check, we compared the expectation value of the bosonic action with the theoretical value obtained using
a supersymmetric Ward identity
\beq
<\kappa S_{\rm boson}> = \left(\frac{3}{2}N_{c}^{2} + N_{c}N_{f}\right)V .
\label{bActionQuiver}
\eeq  Figure~\ref{baction}  
shows a plot of the bosonic action for various values of the soft SUSY breaking coupling $\mu$. In principle we should take the limit $\mu\to 0$ although it should be clear from
the plot that the $\mu$ dependence is in fact rather weak. Up to a constant the expectation value of the bosonic action is just the vacuum energy and the dotted lines in the
plot
show the value that corresponds to $E_{\rm vac}=0$. The red points at the bottom of the figure denote the SUSY preserving case  and
it can be observed that they are consistent with vanishing vacuum energy.  This is to 
be contrasted with the case when $N_f<N_c$  denoted by the blue points which shows a large
deviation from eqn.~\ref{bActionQuiver} and is the first sign that supersymmetry is spontaneously broken in this
case. 
One of clearest signals of supersymmetry breaking can be obtained if one considers the equation of motion for the
auxiliary field.  We expect the susy preserving case to obey
\beq
\frac{1}{N_{c}}  \Tr \left[ \phi(x)\phib(x) \right] = 1.
\label{Tr-ppd}
\eeq 
The red points, corresponding to ($N_{f} > N_{c}$) are consistent with this 
over a wide range of $\mu$. We attribute the small residual deviation
as $\mu\to 0$ to our use of
antiperiodic boundary conditions which inject explicit $\cQ$ susy breaking into the system. 
The simulations with $N_f<N_c$ (blue points) however show a clear signal for spontaneous supersymmetry breaking with the value of
this quantity deviating dramatically from its supersymmetric value even as $\mu\to 0$.  \\
\begin{figure}
\begin{center}
\includegraphics[height=50mm]{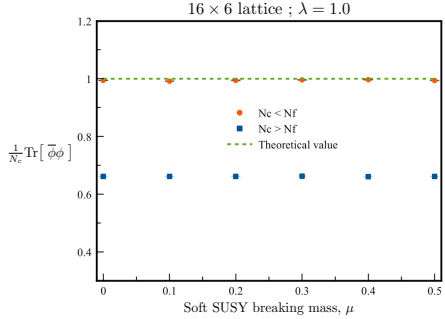}
\caption{$\frac{1}{N_c}{\rm Tr}\phi\phib$ vs $\mu$ for a 't Hooft coupling of $\lambda = 1.0$ on an 16x6 lattice}
\label{Tr-ppd-fig}
\end{center}
\end{figure} 
\begin{figure}
\begin{center}
\includegraphics[height=50mm]{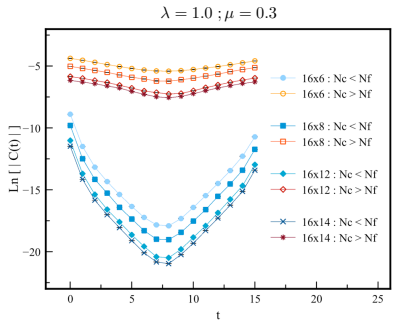}
\caption{Correlation function C(t)  for $\lambda = 1.0$ and $\mu = 0.3$ on various asymmetric lattices}
\label{effb}
\end{center}
\end{figure} 
Finally we turn to our results for a would be Goldstino. We search for this by
computing  the following two point correlation function 
\beq
C(t) = \sum_{x,y}< O^{\prime}(y,t)O(x,0)> 
\label{C(t)}
\eeq where $O^{\prime}(y,t)$ and O(x,0) are fermionic operators given by:
\bea
O(x,0) &=& \psi_{\mu}(x,0)\cU_{\mu}(x,0)\left[\phi(x,0)\phib(x,0) - rI_{N_{c}} \right]
\label{Oprime} \\
O^{\prime}(y,t) &=& \eta(y,t) \left[\phi(y,t)\phib(y,t) - rI_{N_{c}} \right].
\label{O}
\eea 
In figure \ref{effb} we show the
logarithm of this correlator as a function of temporal distance for a range of spatial lattice size, $L=6,8,12$ and 14. The anti-periodic boundary condition is applied along the temporal direction corresponding to T=16 for both $N_f>N_c$ and $N_f<N_c$. Clearly a light state with a strong coupling to the would be Goldstino is seen 
only for the case $N_f<N_c$.


\section{Conclusions}

In this  talk we describe the construction and simulation of a lattice theory of 2d super QCD with gauge group $U(N_c)$ and global $U(N_f)$ flavor symmetry.
The model in question possesses $\cN=(2,2)$ supersymmetry in the continuum limit
while our lattice formulation preserves a single exact supercharge for non zero lattice spacing. It is expected that
the single supersymmetry will be sufficient to ensure that full supersymmetry is regained without fine
tuning in the continuum limit. This
constitutes the first lattice study of a supersymmetric theory containing fields which transform
in both the fundamental and adjoint representations of the gauge group. Our lattice action
also contains a $\cQ$-exact Fayet-Iliopoulos
term which yields a potential for the scalar fields. We see strong signals that the system is Higgsed and
that supersymmetry is spontaneously broken for $N_f<N_c$ and $r>0$.

The lattice  constructions discussed in this paper
generalize \cite{Joseph:2013jya} to three dimensional quiver theories leaving open the
possibility of studying 3D super QCD using lattice simulations.



\end{document}